\documentclass{article}
\usepackage{arxiv}

\usepackage[utf8]{inputenc} 
\usepackage[T1]{fontenc}    
\usepackage{booktabs}       
\usepackage{amsfonts}       
\usepackage{nicefrac}       
\usepackage{microtype}      

\usepackage{graphicx}
\usepackage{caption}
\usepackage{subcaption}

\usepackage{hyperref} 
\hypersetup{
    colorlinks=true,
    linkcolor=blue,
    filecolor=magenta,      
    urlcolor=blue,
}
\usepackage{url}            
\usepackage{amsmath}
\usepackage{amssymb}
\usepackage{amsthm}

\usepackage{verbatim}

\usepackage{changepage} 

\usepackage{color} 

\usepackage{algorithm}
\usepackage{algpseudocode}

\usepackage{url}
\usepackage{hyperref}
\usepackage{xcolor}
\hypersetup{
    colorlinks,
    linkcolor={red!50!black},
    citecolor={blue!50!black},
    urlcolor={blue!80!black}
}

\title{Adaptive data collection for intra-individual studies affected by adherence}

\author{
 Greta Monacelli \\
  School of Computing\\
  Dublin City University, \\
  Insight SFI Research Centre for Data Analytics \\
  \texttt{greta.monacelli2@mail.dcu.ie} \\
   \And
 Lili Zhang \\
  School of Computing\\
  Dublin City University, \\
  Insight SFI Research Centre for Data Analytics \\
     \And
 Winfried Schlee \\
  Department of Psychiatry and Psychotherapy\\
  University of Regensburg \\
     \And
 Berthold Langguth \\
  Department of Psychiatry and Psychotherapy\\
  University of Regensburg \\
     \And
  Tomás E. Ward \\
  School of Computing\\
  Dublin City University, \\
  Insight SFI Research Centre for Data Analytics \\  
  \And
 Thomas B. Murphy \\
  School of Mathematics and Statistics\\
  University College Dublin, \\
  Insight SFI Research Centre for Data Analytics \\
}

\begin{document}
\maketitle

\begin{abstract}
Recently the use of mobile technologies in Ecological Momentary Assessments (EMA) and Interventions (EMI) has made it easier to collect data suitable for intra-individual variability studies in the medical field. 
Nevertheless, especially when self-reports are used during the data collection process, there are difficulties in balancing data quality and the burden placed on the subjects. 
In this paper, we address this problem for a specific EMA setting which aims to submit a demanding task to subjects at high/low values of a self-reported variable. 
We adopt a dynamic approach inspired by control chart methods and design optimization techniques to obtain an EMA triggering mechanism for data collection which takes into account both the individual variability of the self-reported variable and of the adherence rate. 
We test the algorithm in both a simulation setting and with real, large-scale data from a tinnitus longitudinal study. 
A Wilcoxon–Mann-Whitney Rank Sum Test shows that the algorithm tends to have both a higher $F_{1}$ score and utility than a random schedule and a rule-based algorithm with static thresholds, which are the current state-of-the-art approaches. 
In conclusion, the algorithm is proven effective in balancing data quality and the burden placed on the participants, especially, as the analysis performed suggest, in studies where data collection is impacted by adherence. 
\end{abstract}


\section{Introduction}

In recent years, data for intra-individual studies in the medical field have become more easily available thanks to mobile technologies, which also allow to deliver more complex and challenging experimental designs. Intra-individual variability is an important aspect in medical research. For example, it has been used to detect neurological and psychological conditions \cite{macdonald_intra-individual_2006}. Moreover, it can negatively impact medical studies as differences between groups may become less evident, especially for small sample-sized studies \cite{macdonald_intra-individual_2006}. While intra-individual variability is often mentioned in clinical studies, few works specifically focus on it \cite{macdonald_intra-individual_2006}. Recently, more data suitable for intra-individual clinical studies has been collected in recent years through Ecological Momentary Assessments (EMA) and Interventions (EMI) thanks to the use of mobile technologies \cite{de_vries_smartphone-based_2021}. In addition, these outside-the-lab experiments allow for more complex research goals and designs. For example, it is possible to send notifications to the participants based on the recognition of a relevant context such as symptoms severity or physical location \cite{de_vries_smartphone-based_2021, hekler_tutorial_2018, hulme_adaptive_2021}. However, as the complexity of the experiments increases, the logistic challenges to correctly deliver them also increase and efficiency in data collection becomes necessary to reduce the costs of a prolonged experiment such as an EMA or EMI. The term costs may refer to different downsides of the experiment based on the context; examples are the energy consumption of the mobile application \cite{hekler_tutorial_2018} or the burden placed on the subjects during the data collection process.

In particular, EMA and EMI which rely on frequent self-reports and/or lengthy tasks, can be challenging to deliver as they impose significant burdens on participants; in this context, there is a need for methods which reduce the burden on the subjects while preserving data quality. Several studies in the behavioural and psychological field rely on self-reports \cite{de_vries_smartphone-based_2021}. This is especially important in clinical trials where such data collection is described as that gathered by electronic patient reported outcomes or ePRO systems. The type of data collected here may be a more convenient substitute for more objective biological markers which are currently either costly, difficult to deploy or still under active research and validation \cite{gagne_impaired_2020,ahn_computational_2016,mitchell_fmri_2001}. For example, in emerging fields such as computational psychiatry the type of data collected under such approaches constitute the primary source for the development of potentially new biomarkers based on computational modelling \cite{ahn_computational_2016}. Hence, self-reports and data collection "in the wild" in ecologically valid settings are growing increasingly common as these methods provide opportunities to better understand human behaviour, health and disease.  In the clinical setting the potential benefits are improved diagnostics and/or efficacy of therapeutic interventions. Adherence in such a context is an enormous problem for researchers as the approach can place excessive burden placed on the subjects during a longitudinal experiment leading to drop-out and/or poor quality data. All this suggests that methods to balance data quality and the load on the participants are needed. 

In this paper, we develop an algorithm which balances data quality and commitment in a specific EMA setting. The goal of the EMA is to submit a demanding task to subjects at high/low values of a self-reported variable. The researcher fixes a-priori the average desired number of additional tasks to be triggered. Data quality is then measured based on how many tasks are triggered by the algorithm for significantly extreme values. The algorithm is heavily inspired by control chart approaches and adapts to the individual subject based on their past history of self-reports \cite{fricker_comparing_2008}. Moreover, it uses a design optimization method to change the definition of high and low values for the self-reported variable in an adaptive way, based on the expected adherence of the individual subject. Thus, the algorithm is designed to collect data as soon as possible for non-adherent subjects and, in contrast, to burden the adherent subjects only when strictly necessary. The algorithm has been motivated by the needs of the authors to develop a personalised data acquisition schedule for investigating cognitive effects of symptom severity in tinnitus patients. In that clinical trial we require that the participants engage in a relatively lengthy decision making task (a type of two-armed bandit task) when they are as close as possible to the extremes of their self reported symptoms. We require a number of repetitions of this high burden task at these extreme symptom levels. As the participants symptoms fluctuate over time it is required that we use a shorter data capture, in this case a brief self-report on symptoms using an app on the participants' smartphones to track this. Using this data collected several times per day, we must then decide when to trigger the request for participation in the much higher burden decision making task. The non-adaptive strategy options consists on either a random schedule which triggers the task independently from the self-reported symptoms or a rule-based approach based on pre-fixed thresholds. However, we know from prior studies that tinnitus patients present a diversity of self reported symptom ranges \cite{probst_emotional_2016}. Therefore, that the former triggering strategies may not trigger at all for some patients and excessively trigger for others. In this case, we expect to require much more data and better adherence to be sufficiently "lucky" to capture data at the extremes targeted. This reason necessitated the development of the adaptive triggering scheduling algorithm described here. 

The algorithm outperforms, according to our metrics, the state-of-the art approaches - random schedule and rule-based schedule with predetermined thresholds for both simulated and real tinnitus data; moreover, our results suggest that the algorithm is particularly effective when the adherence of the subjects is highly variable. We used the $F_{1}$ scores and the global utility to measure, respectively, the accuracy of the algorithm as a binary classification method and data quality, i.e. its effectiveness in triggering a pre-fixed number of additional tasks to balance the necessities of the researcher and the burden imposed on the subjects. The empirical cumulative distribution functions of these measures were compared pairwise for a random schedule, a rule-based schedule with fixed thresholds and two different adaptive algorithms developed in this paper. In particular, the first algorithm estimates the adherence of the subjects based on prior knowledge on the field, while the second algorithm only utilizes the prior history of the individual subjects to obatin an adaptive estimates of the adherence. The algorithm tends to have higher $F_{1}$ score (p-values $<1.5\times 10^{-3}$ and $<1.2\times 10^{-9}$) and utility (p-values $<3.2\times 10^{-5}$ and $<1.9\times 10^{-13}$) than its competitors for both the simulated and real data. Finally, we find that the p-values obtained when comparing the second adaptive algorithm with the first one are smaller in the real data than in the simulated data; this suggests that the algorithm is particularly suitable to handle high-variability-adherence scenarios such as it is the case for the real data case. 

In conclusion, our algorithm balances the need of collecting sufficient quality data under relatively extreme values of a quantity of interest, as for example tinnitus severity, with that for minimizing burden on the study participants. It can be used for data collection in intra-individual variability studies which rely on self-reports and that in addition place a significant burden on the subjects. We expect the algorithm to be particularly useful when the study is strongly impacted by low adherence rates - a very common situation in EMA studies which rely on self-reports. 

\subsection{Previous work}

Several adaptive methods have been developed to improve data collection in EMA and EMI, but none focuses on the delivery of an additional and demanding task for a predetermined number of times at extreme self-reported values or on the explicit use of the adherence history to improve the algorithm. For example,  \cite{mohan_exploring_2021} and \cite{hekler_tutorial_2018} increase the efficacy of EMI through adaptive rule-based algorithms. \cite{hekler_tutorial_2018} propose a statistical approach to decrease the energy consumption of the mobile application used in the study while maintaining the effective monitoring of a specific heart-condition. Additionally, \cite{thomas_behavioral_2015} and \cite{hulme_adaptive_2021} have developed solutions which deliver just-in-time notifications to the participants through approaches based, respectively, on predetermined rules and hidden Monte Carlo methods.

Previous work from two different fields has been combined to obtain the results in this paper: anomaly --- or outlier --- detection and design optimization. A similar approach has been used by \cite{daiki_automatic_2012} to  search for an optimal significance level in an anomaly detection problem. First, we looked at the anomaly/outlier detection literature to identify extremes values for the quantity of interest. Identifying unusual patterns in the data concerns many different applications and there is a vast literature both in the computational field - anomaly detection \cite{chandola_anomaly_2009} - and in the statistical field -  outliers detection \cite{zimek_survey_2012}. Secondly, we used a design optimization perspective to obtain a more balanced definition of extreme values which takes into account the adherence of the subjects and, thus, the sample size of the data they provide during the experiment. Design optimization methods have been applied in various contexts to balance data quality and experimental costs: examples include clinical trials for drug testing \cite{sylvester_bayesian_1988} or experiments in computational psychiatry \cite{cavagnaro_adaptive_2010}. 

\section{Method}
\label{sec:method}

Let us consider an $N_{d}$ day long experiment. During this time, a participant in the study is asked to self-report a quantity of interest $N_{h}$ times per day through notifications pushed to their mobile phone. The time at which these notification are sent can be either prefixed or randomized. At the end of the experiment, the participant will have received a total of $N = N_{d}N_{h}$ notifications. The data is collected as a time series $\mathbf{x} = \lbrace x_{t}\rbrace_{t =1}^{N}$, where $x_{t} \in [0,1] \cup \lbrace \mathrm{NaN}\rbrace$, $t \in \lbrace  1, \ldots, N\rbrace$ is the discrete time and  $ \mathrm{NaN} $ denotes missing values. 

In this context, we wish to trigger an additional task for \textit{high} and \textit{low} values of the quantity of interest $x_{t}$ for a predetermined number of times $v$. The additional task is assumed to be burdensome or costly for either the participants or the researcher, so that triggering it for every interaction of the subject with the app is not feasible. 

There are two possible immediate solutions to this problem: the first option is a \textit{random algorithm} which randomly selects a subset of the discrete times $R \subset \lbrace 1,\ldots,N\rbrace$ to trigger the task; the second option is a \textit{static algorithm} which triggers the additional task for values of $x_{t}$ above or below predetermined thresholds. Several studies in the behavioural field have implemented either the first or the second options in similar context to the one tackled in this paper \cite{de_vries_smartphone-based_2021}. 

In this paper, we use a control chart approach for this problem. In the anomaly detection and the outlier detection literature this method is usually used to identify observations which are \textit{unlikely} under a probability model, i.e. \textit{outliers}. Nevertheless, it can be applied in the context of this paper by assuming that no outliers or anomalies are present in the data. Indeed, if the data are real-valued and follow a $L^{2}$ distribution, \textit{unlikely} and \textit{extreme} values coincide thanks to the Chebyshev's inequality.

\subsection{Control Chart}
\label{sec:control chart}

Let us first consider the simplified case of a completely adherent participant, so that the time series $\mathbf{x}$ does not contain any missing values ($\mathrm{NaN}$). Assume that $\mathbf{x}$ is an i.i.d. sample from a random variable $X \sim \mathrm{Beta}(\delta,\xi)$, i.e. a Beta distribution of parameters $\delta$ and $\xi$. Let the \textit{triggering starting point} $S\in\lbrace 2,\ldots,N\rbrace$ be the first time point after which the additional task can be triggered. For every $s \in \lbrace S,\ldots,N\rbrace$, consider the unbiased estimators of the expected value and the variance of the sub-sample  $\mathbf{x}_{1 :s}=\lbrace x_{t}\rbrace _{t=1}^{s}$, i.e.
\begin{equation*}
     \hat{\mu}_{s} = \frac{1}{s}\sum_{r=1}^{s}x_{r} \quad \mathrm{and} \quad \hat{\sigma}^{2}_{s} = \frac{1}{(s-1)}\sum_{r=1}^{s}(x_{r}-\hat{\mu}_{s})^{2}.
\end{equation*}
Assume that, for all $ s \in \lbrace S,\ldots,N\rbrace$, $0<{\hat\sigma}_{s-1}^{2}<{\hat\mu}_{s-1}(1-{\hat\mu}_{s-1})$.  Then, the estimates of the parameters $\delta$ and $\xi$  based on the sub-sample $\mathbf{x}_{1 :s}$ obtained through the method of moments are
\begin{equation}
\label{eq:param_est_method of moments}
\hat{\delta}_{s}  = \hat{\mu}_{s-1}\nu_{s-1} \quad \mathrm{and} \quad 
\hat{\xi}_{s}  = (1-\hat{\mu}_{s-1})\nu_{s-1}, 
\end{equation}
where $\nu_{s-1} = \hat{\mu}_{s-1}(1-\hat{\mu}_{s-1})/\hat{\sigma}_{s-1}^{2}- 1$. If  $0<\hat{\sigma}_{s-1}^{2}<\hat{\mu}_{s-1}(1-\hat{\mu}_{s-1})$ does not hold, there may be computational problems. We found that adding two dummy observations to the data, e.g. $0.4$ and $0.6$ solve this issue by slightly increasing the variance of the dataset. 
 
Let $\alpha \in [0,1]$ be the \textit{significance level}. Then, we say that the observation $x_{t}$ is \textit{extreme} if 
\begin{equation}
\label{eq:confidence interval if level alpha - second form}
x_{t} < z_{\frac{\alpha}{2}}(\hat{\delta}_{t-1},\hat{\xi}_{t-1}) \quad \mathit{or} \quad x_{t} > z_{1-\frac{\alpha}{2}}(\hat{\delta}_{t-1},\hat{\xi}_{t-1}),
\end{equation}
where $t\geq S$ and $z_{\beta}(\delta,\xi)$ is the $\beta$-quantile of the Beta distribution with parameters $\delta$ and $\xi$. If Condition \eqref{eq:confidence interval if level alpha - second form} is satisfied at time $t$, then the additional task is triggered. 

\subsection{Selection of the significance level and the triggering starting point }
\label{sec:selection of the triggering starting point and of the significance level}

In Section \ref{sec:control chart}, the criteria to select the triggering starting point $S$ and the significance level $\alpha$ are not specified. In the outliers and anomaly detection literature, both of these design variables are usually chosen by the researcher based on domain knowledge or previous studies \cite{chandola_anomaly_2009}.  In this section, we investigate the choice in the formal framework of design optimization in order to aid the researcher in the final selection \cite{chaloner_bayesian_1995} and following a similar approach to \cite{daiki_automatic_2012}. 

First, we define a utility function that represents the goal of the researcher. We assume that the primary goal of the study is to trigger the additional task on average $v\in \lbrace 1,\ldots,N\rbrace$ times for each participant.

In order to define the utility, let us consider the time series $\mathbf{w}=\lbrace w_{t}\rbrace_{t =S}^{N}$ defined as

\begin{equation*}
w_{t} = \begin{cases} 1 \: \textrm{ if the control chart triggers the additional task at time } t\\
0 \: \mathrm{ otherwise }.
\end{cases}
\end{equation*}

Assume that $\mathbf{w}$ is an i.i.d. sample from a Bernoulli random variable $W \sim \mathrm{Ber}(\alpha)$. Let the random variable $V$ represent the total number of additional tasks triggered during the experiment. Then, $V \sim \mathrm{Bin}(\alpha, N-S+1)$ follows a Binomial distribution of parameters $\alpha$ over $N-S+1$ independent experiments. Thus, the expected value of  $V$ is equal to $\mathrm{E}(V)=(N-S+1)\alpha$. We choose as utility function $U_{1}:\lbrace1,\ldots,N\rbrace\times[0,1] \rightarrow \mathbb{R}$ defined as 
\begin{equation*}
    U_{1}(S,\alpha) = -(\mathrm{E}(V)-v)^{2}.
\end{equation*}
Thus, $U_{1}(S,\alpha)$ associates higher value to experimental designs $(S,\alpha)$ for which the expected number of triggers $\mathrm{E}(V)$ is as close as possible to desired one $v$. 

Let $D=\lbrace(S,\alpha) \in \lbrace 2,\ldots,N\rbrace\times[0,1] \rbrace$, then we aim to find an optimal design $ (S^{\star}, \alpha^{\star})$ such that
\begin{equation}
\label{eq:design optimization problem}
    (S^{\star}, \alpha^{\star}) \in \mathrm{arg}\max_{(S,\alpha) \in D} U_{1}(S,\alpha), 
\end{equation}
The design optimization problem in Equation \eqref{eq:design optimization problem} has as solutions 
\begin{equation}
\label{eq:A^(*)}
    A^{\star} = \Big\lbrace (S^{\star},\alpha^{\star}) \in D \:|\: \alpha^{\star} = \min\Big(0,\max\Big(1, \frac{v}{(N-S^{\star}+1)}\Big)\Big)\Big\rbrace.
\end{equation}
Therefore, experimental designs chosen in the set $A^{\star}$ are all optimal for collecting, on average, $v$ additional tasks. 

\begin{figure}
\begin{center}
  \includegraphics[width=\textwidth]{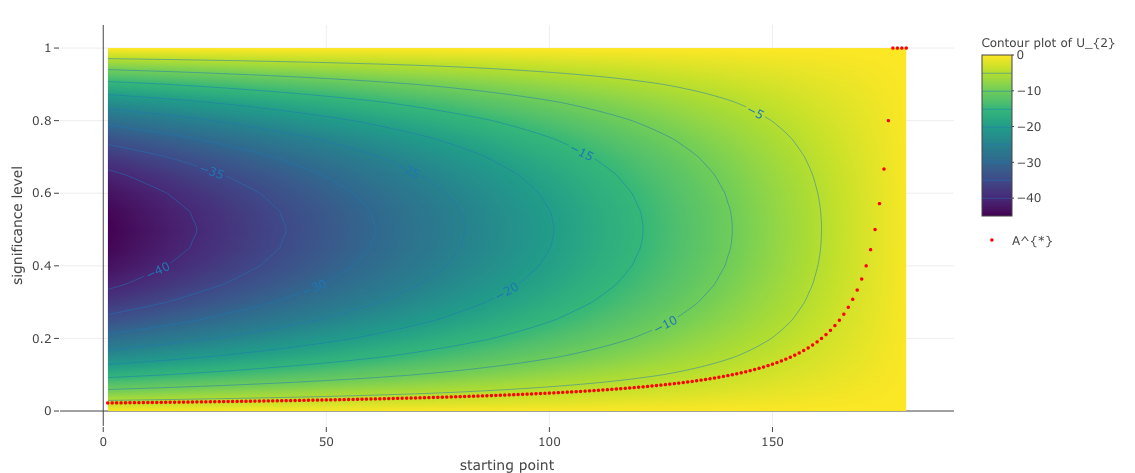}
\caption{The red dots represent the set $A^{\star}$ in Equation \eqref{eq:A^(*)}.  Notice that, as a function of the optimal starting point $S^{\star}$, the optimal significance level $\alpha^{\star}$ increases slowly in the first half of the experiment. The  extension of the function $U_{2}$ in Equation \eqref{eq:U2} into the domain $D' = [1,N] \times [0,1]$ is represented by the contour plot.  }
\label{fig:do_problem}
\end{center}
\end{figure}

Smaller values of the significance level $\alpha$ should be preferred since they result in more significant intra-individual variability at the end of the experiment. If $\alpha^{\star}$ is chosen to be small, then, by the constraint in Equation \eqref{eq:A^(*)}, the starting point $S^{\star}$ is also small. However, while smaller values of $S^{\star}$ can be useful in case of low adherence, they will also likely result in worse data quality since control charts rely on past observations to improve over time. 

We suggest to first fix $S^{\star}$ based on the domain knowledge of adherence in the field and, in any case, to not choose too small a value as it can negatively impact the estimation of the parameters $\hat{\delta}_{s}$ and $\hat{\xi}_{s}$. Once $S^{\star}$ has been fixed, Equation \eqref{eq:A^(*)} gives the optimal significance level $\alpha^{\star}$, i.e.
\begin{equation}
\label{eq:sign level_tot adherence case}
\alpha^{\star} =
\begin{cases} 0 &  \mathrm{ if } \quad N-S^{\star} + 1 \leq 0\\
1 &  \mathrm{ if }\quad 0 \leq N-S^{\star} + 1 \leq v\\
\frac{v}{(N-S^{\star}+1)} & \mathrm{ otherwise }.
\end{cases}
\end{equation}
Plotting the optimal region can further aid the researcher in this selection of the experimental design. For example, in Figure \ref{fig:do_problem} we considered an experiment of length $N=180$ and assumed $v=4, S=6$. In this case, the significance level increases slowly as the starting point increases. This suggests that any starting point $S$ in the first half of the experiment should not impact significantly the final performance of the algorithm for completely adherent subjects. 

Combining Equation \eqref{eq:A^(*)} and the control chart in Section \ref{sec:control chart}, we obtain Algorithm \ref{alg:A1}. Notice that, since the thresholds in the control chart are symmetric with respect to the probability, Algorithm \ref{alg:A1} triggers on average $v/2$ additional tasks for both high and low values. 

\begin{algorithm}[htb]
\caption{}\label{alg:A1}
\begin{algorithmic}[1]
\State Set $N \in \mathbb{N}^{+}$ and $S^{\star}$, $v \in \lbrace 2, \ldots, N\rbrace$. Set $\alpha =  \min\Big(0,\max\Big(1, \frac{v}{(N-S^{\star}+1)}\Big)\Big)$.
\For{$t = 1\ldots,N$} 
    \State collect $x_{t}$
    \If{$t\geq S^{\star}$  and $x_{t}\neq NaN$}
        \State compute $\hat{\delta}_{t-1}$ and $\hat{\xi}_{t-1}$ as defined in Equation \eqref{eq:param_est_method of moments}
        \If{Condition \eqref{eq:confidence interval if level alpha - second form} holds}
            \State trigger the additional task\;
        \EndIf
    \EndIf
\EndFor
\end{algorithmic}
\end{algorithm}

\subsection{Stopping Rule}
\label{sec:stopping rule}

Reducing the variance of $V$ is a second meaningful objective: indeed, once the mean has been fixed, this concentrates the values of $V$ in proximity of $\mathrm{E}(V)$. Recall that $\mathrm{Var}(V)=(N-S+1)\alpha(1-\alpha)$ and define a second expected utility function $U_{2}:\lbrace 1,\ldots,N \rbrace \times [0,1] \longrightarrow \mathbb{R}$ as 
\begin{equation}
\label{eq:U2}
    U_{2}(S,\alpha) = - \mathrm{Var}(V).
\end{equation}
A contour plot of the continuous extension of the function $U_{2}$ into the domain $D' = [1,N] \times [0,1]$ is given in Figure \ref{fig:do_problem}. 

As the variance is only a secondary goal, we consider the design optimization problem for $U_{2}$ constrained to the domain $A^{\star}$, i.e.
\begin{equation}
\label{eq:do_variance}
    (S', \alpha') \in \mathrm{arg}\max_{(S^{\star}, \alpha^{\star}) \in A^{\star}} U_{2}(S^{\star},\alpha^{\star}).
\end{equation}
Then,
\begin{align*}
    U_{2}(S^{\star},\alpha^{\star})& = - (N-S^{\star} + 1)\alpha^{\star}(1-\alpha^{\star})\\
 							&= -  v(1-\alpha^{\star})
\end{align*}
where the second-last step is justified by the constraint $(S^{\star},\alpha^{\star}) \in A^{\star}$. Thus, the optimal solution for the optimization problem in Equation \ref{eq:do_variance} is $(N-v,1)$. This design corresponds to triggering the additional task in the last $v$ interactions of the subject with the app. There are two main problems with this approach: first, in real applications, there is no guarantee that the subject will adhere to the study until the end; second, in order to trigger the additional task on significantly high or low values $x_{t}$ we aim to implement $\alpha$ as small as possible.

In conclusion, reducing the variability of the total number of triggers produced by the algorithm results in a trivial solution for the design optimization problem. Nevertheless, the argument above shows that there is a trade-off between the need for precision --- decreasing the variance of the total number of observations --- and the need for unlikeliness --- keeping the significance level $\alpha$ small. In this paper we assume that both $S^{\star}$ and, subsequently, $\alpha^{\star}$ are chosen to be relatively small. This results in high variance for the final number of triggers. In other words, the algorithm may trigger either too many or too few tasks for certain subjects. Hence, we suggest to add some simple deterministic rules to either force an additional task if there were not enough or stop their triggering if too many have already been triggered. In this paper, we added a stopping rule, and did not allow the algorithm to trigger the additional task more than $10$ times.  

\subsection{Adherence}
\label{sec:adherence}

The selection of the significance level in Section \ref{sec:selection of the triggering starting point and of the significance level} relies on complete adherence from the subjects to the EMA. Indeed, the definition of optimal significance level $\alpha^{\star}$ in Equation \eqref{eq:sign level_tot adherence case} assumes prior knowledge of the final number of samples $N$. Equivalently, missing data are not considered. Nevertheless, this assumption is almost never satisfied in clinical longitudinal studies which rely on self-reports \cite{de_vries_smartphone-based_2021}. 

Assume the dataset has missing data, then the final total number of samples for the quantity of interest is $N'<N$. In this case, Algorithm \ref{alg:A1} triggers on average a lower number of additional tasks than $v$. Indeed, if $N'>v+S-1$, using Equation \eqref{eq:sign level_tot adherence case} we obtain that
\begin{equation*}
    \alpha^{\star} = \frac{v}{N-S+1} < \frac{v}{N'-S+1}. 
\end{equation*}
Thus, the optimal significance level for the missing data scenario should be larger than the one selected assuming complete adherence. 

In order to tackle this problem, we can replace $N$ in Algorithm \ref{alg:A1} with an estimator of $N'$. If a pilot study has been conducted and data are available to estimate the average number of samples per subject, it is possible to use the mean $\overline{N'} = \sum_{j\in B} N'_{j}$, where $B$ is the set of all subjects in the study. In this paper, we used the tinnitus dataset to obtain the estimate $\overline{N'}$ and, in the following, we will always assume that Algorithm \ref{alg:A1} uses it. Nevertheless, this data is not always available. In those cases, it may be difficult to predict accurately the average adherence of the subjects. Moreover, even if this data is available, small sample size, external factors, interventions and/or individual differences may decrease the accuracy of the algorithm. 

\begin{figure}[htb]
\begin{center}
\includegraphics[width = \textwidth]{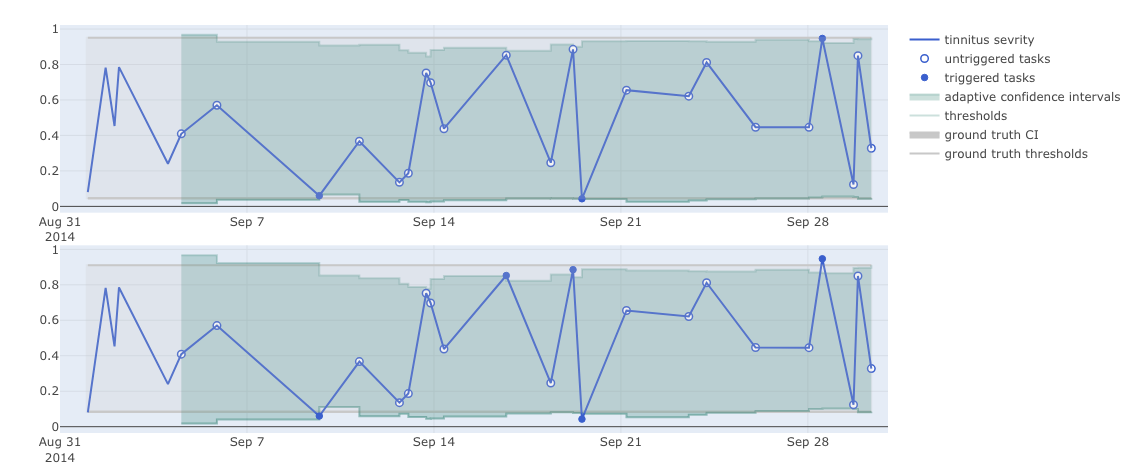}
\caption{Graphical representation of Algorithms \ref{alg:A1} and \ref{alg:A2} in, respectively, the first and the second row. The green area represents the confidence intervals defined adaptively by the algorithm during the collection of the data. The blue line represents the time series of the self-reported tinnitus severity. We set $S^{\star}=6$. Dots colored in white represent not-triggered tasks, while those colored in blue represent triggered ones. The gray area represents the confidence interval of level $\alpha(N)$ as from Equation \eqref{eq:adaptive significance level based on the adherence rate}, i.e. the confidence level computed after all data has been collected. This interval is used to define the ground truth. Therefore, triggered tasks outside this area are defined as true positives, triggers inside are false positive. Non-triggered tasks inside the area are true negatives, non-triggered tasks outside the area are false negatives. Notice that Algorithm \ref{alg:A2} has smaller confidence intervals than Algorithm \ref{alg:A1}, which results in a number of triggered tasks closer to the desired one which had been set to $v=4$ in the simulation.}
\label{fig:reprgraph}
\end{center}
\end{figure}

Another possible approach is to infer a simple adaptive estimate of $N'$ at each time point by considering the history of adherence of the individual subjects and to replace it in Equation \eqref{eq:sign level_tot adherence case}. The advantage of this method is that the estimation of $N'$ does not require prior data.

Let
\begin{equation*}
a_{t} = \begin{cases} 0 \: \mathrm{ if }\: x_{t} = \mathrm{NaN} \\
1 \: \mathrm{ if }\: x_{t} \neq \mathrm{NaN},
\end{cases}
\end{equation*}
so that the time series $\mathbf{a}=\lbrace a_{t}\rbrace_{t =1}^{N}$ represents the adherence of the subjects.

Assume that a participant decides to answer the questionnaire randomly and let the r.v. $A\sim Ber(\chi)$ represent the participant behavior. A success represents an interaction with the app, while a loss represents a missing data. Then, the number of successes $N'$, i.e. the total number of samples we collect from that subject at the end of the experiment, follows a Binomial distribution of parameters $\chi$ and $N$. In particular, 
\begin{equation}
\label{eq:method of moments for adherence}
\mathrm{E}(N') = \chi N.
\end{equation}

We can estimate the probability of success $\chi$ adaptively through the method of moments, thus obtaining
\begin{equation}
\label{eq:adherence rate estimation}
   \hat{\chi}(t) = \frac{\sum_{i=1}^{t}a_{i}}{t} \quad \quad S^{\star}\leq t\leq N'.
\end{equation}
Then, replacing $\chi$ in Equation \eqref{eq:method of moments for adherence} provides the following estimator for the expected value of final samples $N'$: 
\begin{equation}
\label{eq:adaptive stimator of the number of samples}
   \hat{N'}(t) = \hat{\chi}(t) N. 
\end{equation}
Thus, we obtain the following adaptive definition for the significance level, where $\alpha$ is now a function of the time $t$:
\begin{equation}
\label{eq:adaptive significance level based on the adherence rate}
\alpha^{\star}(t) =
\begin{cases} 0 &  \mathrm{ if } \quad \hat{N'}(t)-S^{\star} + 1 \leq 0\\
1 &  \mathrm{ if }\quad 0 \leq \hat{N'}(t)-S^{\star} + 1 \leq v\\
\frac{v}{(\hat{N'}(t)-S^{\star}+1)} & \mathrm{ otherwise }.
\end{cases}
\end{equation}
In conclusion, we obtain Algorithm \ref{alg:A2}.  As suggested in Section \ref{sec:stopping rule}, an additional stopping rule has been added.

\begin{algorithm}[htb]
\caption{}\label{alg:A2}
\begin{algorithmic}[1]
\State Set $N \in \mathbb{N}^{+}$ and $S^{\star}$, $v$, $R \in \lbrace 2, \ldots, N\rbrace$.
\For{$t = 1\ldots,N$} 
    \State collect $x_{t}$
    \If{$t\geq S^{\star}$ and $x_{t}\neq NaN$}
        \State compute $\hat{\delta}_{t-1}$ and $\hat{\xi}_{t-1}$ as defined in Equation \eqref{eq:param_est_method of moments}
        \State compute $\hat{N}'(t)$ and $\alpha^{\star}(t)$ as defined in Equations \eqref{eq:adaptive stimator of the number of samples} and \eqref{eq:adaptive significance level based on the adherence rate}
        \If{Condition \eqref{eq:confidence interval if level alpha - second form} holds and $\sum_{i=1}^{t-1}w_{i} \leq R$}
            \State trigger the additional task
        \EndIf
    \EndIf
\EndFor
\end{algorithmic}
\end{algorithm}

Notice that the definition of extreme values in Algorithm \ref{alg:A2} depends on the adherence of the participant. More adherent participants are expected to have lower values of the adaptive significance level $\alpha^{\star}(t)$ than less adherent ones. Thus, the additional tasks will be triggered for more \textit{significantly} unlikely/extreme values of $x_{t}$ if the participant is adherent. In contrast, the algorithm tends to collect data even if it is not highly significant/extreme from a statistical viewpoint for non-adherent subjects where bad data is better than no data. This characteristic can be both an advantage and a disadvantage. It is an advantage as Algorithm \ref{alg:A2} adapts to the adherence of the subjects as well as to the history of the quantity of interest. On the other hand, it is a disadvantage because, after the collection of the data, this inconsistency should not be overlooked in the subsequent analysis of intra-individual or intra-group differences. 

\section{Results}
\label{sec:Results}

\begin{figure}
\begin{center}
\includegraphics[width = \textwidth]{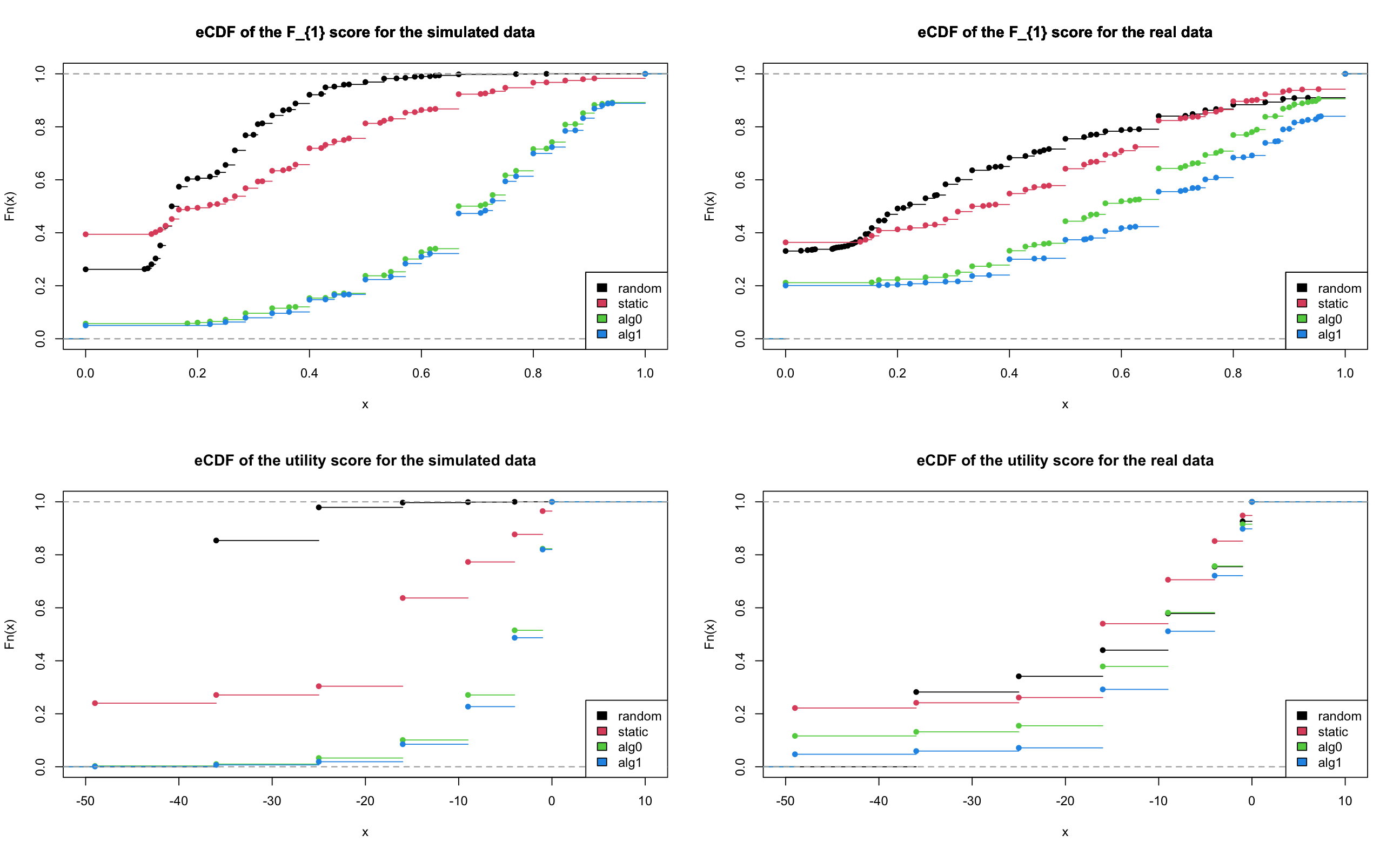}
\caption{Empirical cumulative distribution functions (eCDF) of the $F_{1}$ score and the utility $u_{1}$ in both the simulated --- first column --- and real data --- second column.}
\label{fig:ecdf}
\end{center}
\end{figure}

In this section we compare a random and a static schedule with Algorithms \ref{alg:A1} and \ref{alg:A2} on both a simulation setting and on real data from a tinnitus longitudinal study. The random schedule randomly selects $10$ interactions of the user with the app to trigger the additional task. If the user does not respond to the self-report at the selected times, the additional task trigger shifts to the next interaction of the user with the application until a separate trigger is set. The static algorithm is a rule-based approach with pre-fixed thresholds at $0.15$ and $0.85$. All results in this paper were produced using {\tt{R}} \cite{r_core_team_r_2021}. The Figures were obtained using the {\tt{plotly}} R package \cite{sievert_interactive_2020}.

In both the simulation and the real data analysis, $v=4$ and $N=180$ are fixed. Similarly, the triggering starting point is set to $S^{\star} = 6$. The latter decision was taken by considering the adherence in the real dataset, see Figure \ref{fig:hist-adherence} and trying to balance two contrasting necessities. On one hand, it is important to start collecting the additional task early as the adherence of the subjects decreases exponentially. On the other hand, both Algorithms \ref{alg:A1} and \ref{alg:A2} depend on past observations and are not reliable without previous data. In particular, the method of moments used for the parameter estimation may encounter computational errors and return negative values of the estimated parameters of the Beta distribution if $S^{\star}$ is too small. In practice, we found that $S^{\star}=6$ was a good compromise.

\subsection{Simulated data}
\label{subsec:simulations}

The simulated data were generated from the statistical models used to develop Algorithm \ref{alg:A2}. Therefore, the adherence of the dataset follows a Binomial distribution while the quantity of interest follows a Beta distribution. The Binomial distribution parameters were chosen as $N=180$ and $\chi = 0.19$, where the latter was estimated from the tinnitus dataset. On the other hand, the Beta distribution parameters for the simulated participants are sampled from a uniform distributions in the interval $[0.5, 10]$.

\subsection{Real data}
\label{sec:real data}

In order to investigate the performance of the algorithms in a real scenario, we consider a dataset collected during a tinnitus-focused study
where moment-to-moment tinnitus symptoms were reported using a mobile application called TrackYourTinnitus \cite{probst_emotional_2016}. Since intra-individual fluctuations may be an important characteristic feature of an individual's tinnitus and it is of great importance to investigate the behavior changes while the individual is suffering from tinnitus, this dataset is with significant practical value to test our triggering algorithm.

From the original data, we considered only the \textit{user\_id}, \textit{save}, \textit{save\_date} and \textit{question\_2} columns. These columns collect, respectively, the users identification number, the time and/or date on which the user interacted with the TrackYourTinnitus application and the self-reported tinnitus severity. The tinnitus severity is represented by a value between $0$ and $1$. From this subset of the original data, we delete all rows which have at least a missing value and all duplicates. For some subjects, different tinnitus severity values were reported by the same user at the same identical time; in those cases, we only consider the first-listed of such observations. Since both Algorithm \ref{alg:A1} and \ref{alg:A2} require a pre-fixed length of the experiment, we further restrict our attention to only the $30$ days after the first interaction of each participant with the application. For most of the participants ($77\%$, or $2130/2752$) there is no information loss as they interacted with the application only within this time-frame. Finally, as the starting point has been fixed to $S^{\star}=6$, we do not consider users which interacted less than $6$ times with the application. Notice that, after cleaning the data, only around $33\%$, ($911/2752$) of the initial users remain. 

\subsection{Statistical analysis}

\begin{table}[htb]
\begin{center}
\caption{P-values of the Wilcoxon–Mann–Whitney Rank Sum Test for the simulated data. In the first row, we find the p-values relative to the distribution of the $F_{1}$ score. The notation \textit{algorithm} $>$ \textit{algorithm'} denotes that the distribution of the $F_{1}$ score of \textit{algorithm} is stochastically greater than the one of \textit{algorithm'}. Equivalently, \textit{algorithm} tends to have higher $F_{1}$ scores than \textit{algorithm'}. The notation \textit{algorithm} $>$ \textit{algorithm'} represents the alternative hypothesis of the test, which is rejected with the p-value listed in the Table. The same explanation applies to the second row of the table, which is relative to the distribution of the global utility $u_{1}$.}
\label{tab:wlc_sim}
\begin{small}
\begin{tabular}{lllllll}
\hline
                   & static $>$ random & alg1 $>$ random &
 alg1 $>$ static & alg2 $>$ random & alg2 $>$ static & 
 alg2 $>$ alg1 \\ 
 \hline
 $F_{1}$ & $8.9\cdot 10^{-12}$ & $	
2.8\cdot 10^{-149}$ & $3.8\cdot 10^{-125}$ & $	
1.4\cdot 10^{-154}$ & $5.6\cdot 10^{-131}$ & $1.5\cdot 10^{-3}$\\  
 $u_{1}$ & $5.0\cdot 10^{-89}$ & $5.8\cdot 10^{-165}$ & $4.5\cdot 10^{-113}$  & $1.6\cdot 10^{-165}$ & $9.1\cdot 10^{-121}$ & $3.2\cdot 10^{-5}$ \\
\end{tabular}
\end{small}
\end{center}
\end{table}

\begin{table}[htb]
\begin{center}
\caption{P-values of the Wilcoxon–Mann–Whitney Rank Sum Test for the real data. See Table \ref{tab:wlc_sim} for more information.}
\label{tab:wlc_real}
\begin{small}
\begin{tabular}{lllllll}
 \hline
                   & static $>$ random & alg1 $>$ random &
 alg1 $>$ static & alg2 $>$ random & alg2 $>$ static & 
 alg2 $>$ alg1 \\ 
 \hline
 $F_{1}$ & $3.4\cdot 10^{-4}$ & $	
7.6\cdot 10^{-29}$ & $1.4\cdot 10^{-30}$ & $	
4.1\cdot 10^{-46}$ & $5.6\cdot 10^{-42}$ & $1.2\cdot 10^{-9}$\\  
 $u_{1}$ & $1$ & $2.4\cdot 10^{-2}$ & $2.1\cdot 10^{-21}$  & $4.5\cdot 10^{-16}$ & $3.3\cdot 10^{-46}$ & $1.9\cdot 10^{-13}$ \\
\end{tabular}
\end{small}
\end{center}
\end{table}

In order to compare the performances of the four algorithms, both the distribution of the $F_{1}$ scores \cite{sokolova_systematic_2009} and the estimated utility $u_{1} := (\sum_{i=1}^{N}w_{i}-v)^{2}$ were computed. The $F_{1}$ score measures the accuracy of the algorithm against the ground truth, which is defined based on Equations \eqref{eq:confidence interval if level alpha - second form} and \eqref{eq:sign level_tot adherence case}. This implied that the ground truth is tailored to the individual subject and varies across individuals (see Figure \ref{fig:reprgraph}). On the other hand, the utility $u_{1}$ measures how effectively the algorithm achieves the goals set by the design optimization problem. The empirical cumulative distribution functions (eCDF) of these quantities for both the simulated data and the real data are shown in Figure \ref{fig:ecdf}. Moreover, we compare the eCDFs pairwise through the Wilcoxon–Mann–Whitney Rank Sum Test \cite{brunner_rank_2018} using the \textit{wilcoxon.test} function in R. Notice that the computation of the $F_{1}$ score produced some $NaN$, which were omitted in the statistical test.  The results are listed in Tables \ref{tab:wlc_sim} and \ref{tab:wlc_real}. 

\begin{figure}
\begin{center}
\includegraphics[width=\textwidth]{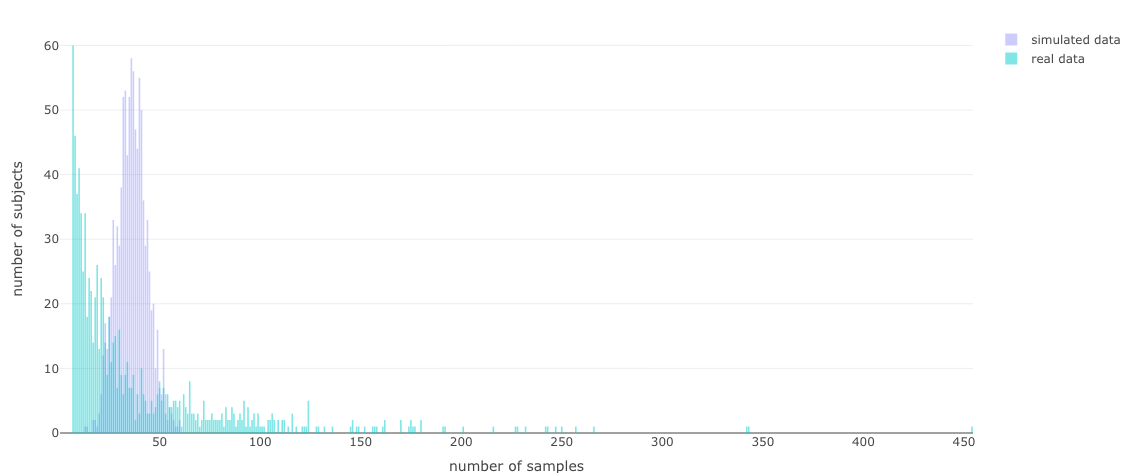}
\caption{Comparison between the total number of observations per subject in the real data and the simulated data set. The parameters used for the simulation were estimated through the method of moments from the real data. Nevertheless, the histograms are not similar in shape. We conclude that adherence in the real data is not well described by a Binomial distribution. Some participants have a higher number of interaction than expected as they interacted with the application more than $6$ times per day. In those cases, all samples are considered to estimate the parameters of the Beta distribution and an additional task can be triggered at those time-points. However, only the first $6$ samples per day are considered for the adherence rate estimation in Equation \eqref{eq:adherence rate estimation}.}
\label{fig:hist-adherence}
\end{center}
\end{figure}

We find that, for both the simulated and the real data, Algorithm \ref{alg:A2} tends to have larger $F_{1}$ scores and utility values $u_{1}$ than all the other options. Moreover, in the comparison between Algorithms \ref{alg:A2} and \ref{alg:A1} we obtain smaller p-values in the real case compared to the simulated one, see the last columns of Tables \ref{tab:wlc_sim} and \ref{tab:wlc_real}. 
This discrepancy may be caused by the greater variability of adherence in the real data in comparison to the simulated data, see Figure \ref{fig:hist-adherence}. This suggests that the additional complexity of Algorithm \ref{alg:A2} compared to Algorithm \ref{alg:A1} is more "useful" for studies where there is a high variability in the adherence of the subjects.  

\section{Discussion}

In conclusion, we have presented a valid alternative to the current state-of-art sampling methods for adaptive EMA based on extreme self-reports. Algorithms \ref{alg:A1} and \ref{alg:A2} have been proven effective to collect data through an additional and burdensome task based on high or low values of a less intrusive self-report. Indeed, they have both outperformed a static and a random triggering approaches in a simulated scenario. The same results are confirmed in the real case scenario of tinnitus severity data. Moreover, the analysis suggest that Algorithm \ref{alg:A2} is slightly more effective than Algorithm \ref{alg:A1} in increasing data quality (obtaining a desired number of additional tasks for a higher value of $u_{1}$) while preserving precision (similar $F_{1}$ scores), especially for studies with high variability in adherence. This is especially important as many fields, such as the tinnitus research area, rely on self-reports for which adherence is a challenging problem. 

Algorithms \ref{alg:A1} and \ref{alg:A2} can be easily adapted to various contexts, but still rely on strong statistical assumptions which may not hold in real scenarios. It is possible to generalize these algorithms to other distributions: arguments in the paper remain valid, with adequate changes to the notation, if the quantity of interest $X$ follows an absolutely continuous distribution. Nevertheless, it is still necessary to have easy access to the quantile of the distribution in order to effectively deploy the algorithm.  On the other hand, we assume that the distribution of the quantity of interest $X$ is known a-priori. In addition, the quantity of interest and the adherence of the subjects are assumed independent and identically distributed even though time is often an important factor in longitudinal studies. Despite the latter limitations, it can be easily adapted to the necessity of the researcher and we have demonstrated its effectiveness through simulations and large scale real world analysis. 

\subsection{Future work}

There are many possible ways to further develop the arguments in this paper. Using statistical models which consider time for both the quantity of interest and the adherence such as autoregressive models could improve the performance of the algorithm in real case scenarios \cite{fricker_comparing_2008}. Moreover, more than one quantity of interest is often collected during a longitudinal study \cite{probst_emotional_2016} and there may be an interest in developing a multivariate version of the algorithms \cite{chandola_anomaly_2009}. In particular, including objective measures, such as measurements from wearable sensors, may increase the effectiveness of the algorithms \cite{macdonald_intra-individual_2006}. 
In this direction, hierarchical methods could inspire future work as they have been successfully applied in the field of anomaly detection \cite{sottas_athlete_2011}.
Finally, the algorithm developed in this paper provides an adaptive sampling schedule for deployment of a secondary burden-heavy task contingent on data from a lighter, more frequent data collection task. Nevertheless, it does not address the sampling scheme for the lighter task. An adaptive sampling scheme which predicts the unlikely states and sends a notification to the participants only when needed could further increase the balance between data quality and the burden placed on the subjects \cite{hulme_adaptive_2021}. 

\subsection*{Supplementary material}
The code and the simulated data are available from the corresponding author or on our lab GitHub repository (\href{https://github.com/AI-for-Better-Living/adc-tinnitus}{https://github.com/AI-for-Better-Living/adc-tinnitus}). Due to privacy and ethical reasons, the data from the TrackYourTinnitus  application are not  publicly available. They are available on request (winfried.schlee@ieee.org).

\subsection*{Acknowledgements}
This research is supported by Science Foundation Ireland (SFI) under Grant Number SFI/12/RC/2289\_P2, co-funded by the European Regional Development Fund.
\vspace*{1pc}

\noindent {\bf{Conflict of Interest}}

\noindent {\it{The authors have declared no conflict of interest.}}

\bibliographystyle{unsrt}  
\bibliography{references.bib}

\end{document}